\documentclass[aps,prb,amsmath,amssymb,twocolumn,superscriptaddress,longbibliography]{revtex4-1}
\usepackage{graphicx}
\usepackage{dcolumn}
\usepackage{bm}
\usepackage{rotating}
\usepackage{booktabs}
\usepackage{siunitx}
\usepackage{multirow}
\usepackage{afterpage}
\usepackage[table]{xcolor}
\usepackage[english]{babel}
\usepackage{amssymb} 
\usepackage{amsmath} 
\usepackage[utf8]{inputenc} 
\usepackage{longtable}
\usepackage{placeins}

\newcommand{\be}{\begin{equation}}
\newcommand{\ee}{\end{equation}}
\newcommand{\bea}{\begin{eqnarray}}
\newcommand{\eea}{\end{eqnarray}}
\newcommand{\nn} {\nonumber}

\newcommand{\Tr}{ {\rm Tr} \, }

\def\a{\alpha}

\def\G{\Gamma}
\def\d{\delta}
\def\D{\Delta}

\def\ve{\varepsilon}

\def\S{\Sigma}

\def\w{\omega}

\def\bra{\langle}
\def\ket{\rangle}

\def\xc{{\rm xc}}
\def\x{{\rm x}}

\def\Tr{{\rm Tr}\,}

\begin{document}
\widetext 

\title{Electronic structure of TiSe$_2$ from a quasi-self-consistent $G_0W_0$ approach}
\author{Maria Hellgren}
\affiliation{Sorbonne Universit\'e, Mus\'eum National d'Histoire Naturelle, UMR CNRS 7590, Institut de Min\'eralogie, de Physique des Mat\'eriaux et de Cosmochimie (IMPMC), 4 place Jussieu, 75005 Paris, France}
\author{Lucas Baguet}
\affiliation{Sorbonne Universit\'e, Mus\'eum National d'Histoire Naturelle, UMR CNRS 7590, Institut de Min\'eralogie, de Physique des Mat\'eriaux et de Cosmochimie (IMPMC), 4 place Jussieu, 75005 Paris, France}
\author{Matteo Calandra}
\affiliation{Dipartimento di Fisica, University of Trento, Via Sommarive 14, 38123 Povo, Italy}
\affiliation{Sorbonne Universit\'es, CNRS, Institut des Nanosciences de Paris, UMR7588, 75252, Paris, France}
\author{Francesco Mauri}
\affiliation{Dipartimento di Fisica, Universita di Roma La Sapienza, Piazzale Aldo Moro 5, I-00185 Roma, Italy}
\author{Ludger Wirtz}
\affiliation{Department of Physics and Materials Science, University of Luxembourg, 162a avenue de la Fa\"{\i}encerie, L-1511 Luxembourg, Luxembourg}
\date{\today}
\begin{abstract}
In a previous work it was shown that the inclusion of exact exchange is essential for a first principles description of both the electronic- and the vibrational properties of TiSe$_2$, M. Hellgren et al. [Phys. Rev. Lett. 119, 176401 (2017)]. The $GW$ approximation provides a parameter-free description of screened exchange but is usually employed perturbatively ($G_0W_0$) making results more or less dependent on the starting point. In this work, we develop a quasi-self-consistent extension of $G_0W_0$ based on the random phase approximation (RPA) and the optimized effective potential of hybrid density functional theory. This approach generates an optimal $G_0W_0$ starting-point and a hybrid exchange parameter consistent with the RPA. While self-consistency plays a minor role for systems such as Ar, BN and ScN, it is shown to be crucial for TiS$_2$ and TiSe$_2$. We find the high-temperature phase of TiSe$_2$ to be a semi-metal with a band structure in good agreement with experiment. Furthermore, the optimized hybrid functional agrees well with our previous estimate and therefore accurately reproduces the low-temperature charge density wave phase. 
\end{abstract}
\maketitle
\section{Introduction}
TiSe$_2$ is a layered quasi-two-dimensional material that undergoes an unconventional charge density wave (CDW) transition below 200 K. The apparent interplay between the CDW and superconductivity at finite pressure or doping\cite{PhysRevLett.103.236401,morosan} has lead to numerous studies over the past years aiming to understand the driving mechanism behind the CDW. Nevertheless, the relative role played by excitonic effects and electron-phonon coupling is still debated. Experimentally, strong signatures are observed in both  vibrational\cite{diSalvo76,PhysRevB.16.3628,PhysRevLett.107.266401,PhysRevLett.91.136402} and angle-resolved photoemission spectra (ARPES),\cite{PhysRevB.61.16213,PhysRevLett.88.226402,PhysRevB.65.235101,PhysRevLett.99.146403,PhysRevB.79.045116,chiang-arpes,aebi19} and some studies point to soft electronic modes.\cite{Kogar1314} 

First-principle calculations should be able to explain the exact mechanism of the CDW transition. However, numerically tractable approaches such as the local density approximation (LDA) or generalized gradient approximations (GGAs) within density functional theory (DFT) fail to give a complete picture.\cite{calandra11,PhysRevLett.107.266401,bianco2015} A dramatic improvement is found when including a fraction of Hartree-Fock (HF) exchange via the hybrid functionals.\cite{chenhyb2015,hse06cdw,hellgrenbaima17} With a result similar to the DFT$+U$ approach,\cite{LDAU,bianco2015} the Ti-$d$ levels are then well described. In addition, the hybrid functionals contain the long-range Coulomb interaction which was shown to be crucial to induce the CDW phase.\cite{hellgrenbaima17} 
This fact suggests that strong electron-hole coupling is at play and that an excitonic transition could be of importance.\cite{cnt2018,exchyb2018} On the other hand, it was also found that the standard medium-range hybrid functional already gives a quantitatively reasonable agreement between theory and experiment. However, the results also showed to be strongly dependent on the hybrid parameters, making it still uncertain whether a parametrization optimized on a test-set of standard semi-conductors is adequate. 

The $GW$ method is computationally more expensive but provides a parameter-free and physical description of screened exchange. The bare Coulomb interaction is replaced by the screened Coulomb interaction, $W$, which is determined by the linear density response function approximated at the Hartree level, i.e., the random phase approximation (RPA).\cite{hedin,PhysRevB.57.2108} The $GW$ approximation for the self-energy is known to produce accurate band-gaps on a wide range of systems.\cite{strinati80,hyblouie,ferdirevgw,luciarevgw} It is, however, almost always employed perturbatively ($G_0W_0$), on top of a DFT Kohn-Sham (KS) band structure, assuming that the KS electronic structure is close enough to the final result. Other variants that bring results closer to self-consistency have also been developed.\cite{PhysRevB.74.045102,PhysRevLett.96.226402,PhysRevB.75.235102,atalla} An alternative to the fully self-consistent $GW$ scheme is to look for the optimal KS starting-point via the Sham-Schl\"uter equation.\cite{lss8812,lss882,rubiooep} The resulting KS potential produces a density similar to the $GW$ density and is known as the RPA potential.\cite{lsshellgren,hellgrenrohr} The KS RPA band structure can be shown to provide a consistent starting-point for $G_0W_0$.\cite{gonze,kresseoep} 

A high-level calculation of the electronic band structure of TiSe$_2$ in the high-$T$ phase would be valuable. While transport experiments all predict a semi-metallic behaviour some ARPES measurements have found a gap.\cite{PhysRevB.61.16213,PhysRevLett.88.226402} The latter scenario was supported by the first $G_0W_0$ calculation and interpreted as an excitonic gap.\cite{cazzaniga2012} In this work, we will re-examine how $G_0W_0$ performs on TiSe$_2$ by first showing that it is a case sensitive to exchange in the starting-point. As a fully self-consistent calculation is out of reach we develop a quasi-self-consistent approach that exploits the local hybrid potential as an approximation to the local RPA potential. In this way, we produce a theoretically justified $G_0W_0$ solution that approximates the RPA solution. At the same time we generate an RPA-optimized hybrid functional that is used to study the CDW phase.

The paper is organized as follows. In Sec. II we start by reviewing the $GW$ formalism and the RPA as a self-consistent way to do perturbative $G_0W_0$. We then introduce a hybrid functional approach based on the optimized effective potential. Using this potential we then develop a quasi-self-consistent $G_0W_0$ scheme and compare it to variants introduced by others. In Sec. III we present numerical results for Ar, BN, ScN, TiS$_2$ and TiSe$_2$. We also use the RPA optimized hybrid functional to study the CDW phase of TiSe$_2$. 
Finally, in Sec. IV we present our conclusions. 

\section{Screened exchange from $GW$} 
We will focus on studying the performance of the $GW$ approximation in describing the band structure of the high-$T$ phase of TiSe$_2$. The results turn out to be strongly dependent on which approximate $GW$ scheme is used. In this section we, therefore, start by reviewing the different ways to solve the $GW$ equations and discuss the connections between $GW$, RPA, COHSEX (COulomb Hole Screened EXchange) and hybrid functionals. This will allow us to finally motivate a quasi-self-consistent $G_0W_0$ approach based on the local hybrid potential.
\subsection{The $GW$ approximation}
We define the self-energy as the nonlocal frequency dependent potential $\S$ that contains all 
the many-body effect beyond the Hartree (H) approximation. To first order, in an expansion in terms of the Green's function, $G$, and the Coulomb interaction, $v$, $\S$ is just the static but nonlocal Fock term of the HF approximation,
\be
\S^{\rm HF}=iGv.
\ee 
By replacing the bare Coulomb interaction in the Fock term with the dynamically screened Coulomb interaction, $W$, we obtain the self-energy within the $GW$ approximation 
\be
\S=iGW.
\ee

The screened interaction within the $GW$ approximation is approximated at the time-dependent Hartree level for which the irreducible polarizability, $P$, is approximated with $P_0$, i.e., to zeroth order in the explicit dependence on the Coulomb interaction. We thus have
\be
W=v+vP_0W, \,\,\,\,P_0=-iGG.
\ee
From Dyson's equation,
\begin{equation}
G=G_{\rm H}+G_{\rm H}\S[G]G,
\label{dysonH}
\ee
we then have access to the many-body quasi-particle spectrum contained in $G$. 

It can further be shown that the $GW$ approximation is a  $\Phi$-derivable approximation\cite{bk1,bk2} that obeys physical conservation laws and has an underlying action functional. An example of such an action functional is the Klein functional\cite{klein}
\bea
Y_{\rm K}&=&-i\Phi[G]-U_{\rm H}+\,i\Tr [GG_{\rm H}^{-1}-1+\ln (-G^{-1})],
\label{klein}
\eea
where $U_{\rm H}$ is the Hartree energy. With the choice
\be
\Phi[G]=\frac{1}{2}{\rm Tr}\{\ln [1+ivGG]\}
\ee
it is easy to see that $Y_{\rm K}$ is stationary when $G$ obeys Dyson's equation (Eq.~(\ref{dysonH})), and the self-energy is equal to 
\begin{equation}
\S=\frac{\d \Phi}{\d G}=iGW.
\ee

At the stationary point the Klein functional is equal to
the $GW$ total energy as obtained from the standard non-variational Galitskii-Migdal energy expression.\cite{gm} 

Instead of using the Hartree approximation as the zeroth order approximation for $G$ one can start from the DFT KS system. The Dyson's equation can then be re-written in terms of the single-particle KS Green's function, $G_s$, and the exchange-correlation (xc) part of the local KS potential 
\be
G=G_s+G_s[\S[G]-v_{\rm xc}]G.
\label{dysonxc}
\ee

The diagonal of $G$, i.e. the density, is already exactly described by $G_s$. In this way, $G_s$ can be assumed to be 'close' to $G$, justifying a perturbative treatment of Eq.~(\ref{dysonxc}), and thus circumventing the full solution to the numerically challenging Dyson's equation. By writing Eq.~(\ref{dysonxc}) in the basis of KS orbitals and keeping only the diagonal terms we can write the quasi-particle equation as\cite{strinati82,hyblouie}
\be
E_k=\ve_{k}+\bra k|\S_s(E_k)-v_{\rm xc}| k \ket
\label{gw_dft}
\ee 
where $k$ refers to the Bloch orbital index. The subscript $s$ on the self-energy signifies that it is evaluated with $G_s$. 
The energy dependence of $\S_s$ can either be treated to zeroth order, i.e. $E_k=\ve_k$, where $\ve_k$ is the KS eigenvalue, or to first order in a Taylor expansion around $\ve_k$. The latter implies that a renormalization factor 
\be
Z_{k}=\left[1-\left.\frac{\partial \Re\S_{s}}{\partial \w}\right|_{\w=\ve_{k}}\right]^{-1}
\ee
should be multiplied in the following way
\be
E_k=\ve_{k}+Z_{k}\bra k|\S_s(\ve_k)-v_{\rm xc}| k \ket.
\label{gw_dft_0}
\ee 
This $G_0W_0$ correction, starting from PBE or LDA, is the most common $GW$ approach to determine the band structure. The justification of this approach relies, however, on the assumption that PBE or LDA orbitals are similar to the true quasi-particle orbitals. The renormalization factors are usually incorporated
but it can be argued that these should be omitted.\cite{gonze} The arguments are based on the connection between $GW$ and the RPA for the total energy,\cite{casida,lsshellgren,caruso_bond_2013} as we will now discuss. 

Let us go back to the Klein energy functional (Eq.~(\ref{klein})) and keep the $\Phi$-functional in the $GW$ approximation. From now on we will add superscripts ($\Phi^{GW},\S^{GW}$) as we focus only on this approximation. If we replace the interacting $G$, in every term, with a non-interacting $G_s$ we can, after a few manipulations, write Eq.~(\ref{klein}) as
\be
Y^{GW}_K[G_s]=-i\Phi^{GW}[G_s]+T_s[n]+U_{\rm H} + U_{\rm ext}
\label{kleinrpa}
\ee
where $T_s$ is the kinetic energy of non-interacting KS electrons and $U_{\rm ext}$ is the external potential energy. It is easy to see that $\Phi^{GW}[G_s]$ is exactly the same functional 
as the xc energy of the RPA energy functional
\be
E^{\rm RPA}_{\rm xc}\equiv -i\Phi^{ GW}[G_s]=-\frac{i}{2}{\rm Tr}\{\ln [1+ivG_sG_s]\}.
\ee
Eq.~(\ref{kleinrpa}) is thus nothing but the RPA total energy, i.e., $Y^{GW}_K[G_s]=E^{\rm RPA}$.\cite{lsshellgren,caruso_bond_2013,PhysRevB.91.165110}

The RPA energy functional can be shown to have a minimum when varied with respect to the total KS potential $V_s=v_{\rm ext}+v_{\rm H}+v_\xc$. Such a variation can be carried out via the functional $G_s[V_s]$. At the minimum $v_\xc=v^{\rm RPA}_\xc$ obeys the so-called linearized Sham-Schl\"uter (LSS) equation
\be
\int\!d2\, \chi_s(1,2) v^{\rm RPA}_{\xc}(2)=\int\!d2d3\,\Lambda_s(3,2;1) \S_s^{ GW}(2,3)
\label{lss}
\ee
where $\Lambda_s(3,2;1)=-iG_s(3,1)G_s(1,2)$ and $\chi_s(2,1)=-iG_s(2,1)G_s(1,2)$.\cite{lss88} The LSS equation can also be derived from the condition that the $GW$ density and the KS RPA density, i.e. the diagonals of $G$ and $G_s$, are the same to first order when expanding Dyson's equation (Eq.~(\ref{dysonxc})).

As the RPA potential is a local KS potential it does not reproduce the fundamental gap.\cite{lss8812,lss882,rubiooep} One can, however, still calculate the gap, $E_g$, corresponding to the RPA functional by taking the derivative of the energy functional with respect to particle number $N$. One finds 
\be
E_g=I-A=\left.\frac{\partial E^{\rm RPA}}{\partial N}\right|_{+}-\left.\frac{\partial E^{\rm RPA}}{\partial N}\right|_{-}.
\ee
Evaluating the derivative on the right hand side '+', i.e., the negative of the ionization energy 
\be
-I=\ve_{v}+\bra v|\S_s^{GW}(\ve_{v})-v^{\rm RPA}_{\rm xc}| v \ket
\label{ion}
\ee 
and the derivative on the left hand side '-', i.e., the negative of the affinity 
\be
-A=\ve_{c}+\bra c|\S_s^{GW}(\ve_{c})-v^{\rm RPA}_{\rm xc}| c \ket
\label{aff}
\ee 
we can write
\be
E_g=E_g^{\rm KS}+\D_\xc
\label{rpadisc}
\ee
where $E_g^{\rm KS}$ is the KS gap and
\be
\D_\xc=\bra c|\S_s^{GW}(\ve_{c})-v^{\rm RPA}_{\rm xc}| c \ket-\bra v|\S_s^{GW}(\ve_{v})-v^{\rm RPA}_{\rm xc}| v \ket.
\ee 
 To derive these expressions Eq.~(\ref{lss}) has to be used. The quantity $\D_\xc$ equals what is called the derivative discontinuity 
 within DFT.\cite{pplb,gunnshon,hellgren12,hellgren13} 
 
 It is now clear that the gap obtained from the RPA functional is nothing but the $G_0W_0$ correction of Eq.~(\ref{gw_dft_0}), without the $Z_k$ factor, evaluated with the RPA potential. The RPA potential can thus be seen as an optimal KS starting point for $G_0W_0$, which produces a gap equal to the gap extracted from the RPA functional.\cite{gonze} It has been shown on a number of semiconductors\cite{kresseoep} that using the RPA potential for a $G_0W_0$ calculation brings gaps in closer agreement with self-consistent $GW$ approaches.\cite{PhysRevB.75.235102}
 
By expanding the $GW$ quasiparticle energy around the zeroth order RPA KS energy and using Eq.~(\ref{lss}) the expressions in Eqs.~(\ref{ion})-(\ref{aff}) are easily extended to the whole band structure,\cite{gonze}
\be
E_k^{\rm RPA}=\ve_{k}+\bra k|\S^{GW}_s(\ve_{k})-v^{\rm RPA}_{\rm xc}| k \ket.
\label{gw1}
\ee 

To conclude we have reviewed how it is possible to calculate gaps and even the full band structure from the RPA and that this corresponds to the perturbative $G_0W_0$ result evaluated with the local RPA potential. These are well-known results that we will base the following discussions on. 
\subsection{Hybrid functionals and the COHSEX approximation}
We will now turn to the simpler COHSEX and hybrid functionals which are often used as cheaper but self-consistent alternatives to the $GW$ approach. 

The frequency dependence of the $GW$ self-energy 
allows for the description of many-body effects such as quasi-particle lifetimes and satellite spectra but severely complicates a fully self-consistent solution. 
Often one is, however, only interested in the quasiparticle excitation energy for which the nonlocality of the self-energy plays the most important role. It is then motivated to approximate $\S$ by ignoring dynamical effects in $W$. This implies setting
\bea
W_{\rm static}&=&v+vP_0(\w=0)W_{\rm static},
\eea
and results in the so-called COHSEX approximation 
\be
\S^{\rm COHSEX}=iGW_{\rm static}+\frac{1}{2}W_p^d,
\ee
where $W_p^d={\rm Diag}\left[vP_0(\w=0) W_{\rm static}\right]$ is a local Coulomb-hole potential and the first term is a nonlocal statically screened exchange operator. 
The COHSEX approximation can easily be solved self-consistently but can still be numerically demanding since it requires the generation and summation over all conduction bands. A more drastic approach that avoids the inclusion of unoccupied bands is to keep the bare Coulomb interaction as in the HF approximation but scale it down with a constant $\a$. If we then add a compensating fraction of the local PBE exchange and a local PBE correlation term we get the so-called hybrid functionals 
\be
\S^{\rm HYB,\a}=\a\S^{\rm HF}+(1-\a)v^{\rm PBE}_{\x}+v^{\rm PBE}_{\rm c}.
\ee
These functionals are structurally similar to COHSEX but not more demanding than a HF calculation. One of the drawbacks is that a free parameter is introduced. A fraction 25\% (PBE0) has shown to be reasonable in many molecular systems. In the HSE06 functional a second parameter, $\mu=0.2\,{\rm \AA}^{-1}$, that controls the range of the Coulomb interaction is introduced.\cite{hse03} In this way, it is possible to get a good description of many semiconductors as well. 

Although often used in a DFT context the hybrid functionals are almost always treated like the HF approximation, that is, by minimizing the energy with respect to orbitals that are generated by the nonlocal Fock potential. 
In this work we will instead use the optimized effective potential method\cite{oep} and minimize the hybrid energy with respect to a local KS potential. The local KS potential corresponding to HF has been evaluated for solids before and is know as the exact-exchange (EXX) potential.\cite{gorlingexx,engel} The local hybrid potential is given by the sum of the local potentials derived from the PBE terms and a local exchange potential obtained from an equation similar to the LSS equation (Eq.~(\ref{lss})) but with $\S_s^{GW}$ replaced by the scaled HF self-energy. We have 
\be
v_\xc^{\rm hyb,\a}=v^{\a}_{\x}+(1-\a)v^{\rm PBE}_{\x}+v^{\rm PBE}_{\rm c}
\label{lhyb}
\ee
where
\be
\int\!d2\, \chi_s(1,2) v^{\a}_{\x}(2)=\a\int\!d2d3\,\Lambda_s(3,2;1) \S_s^{\rm HF}(2,3).
\label{exxlss}
\ee
The potential $v^{\a}_{\x}$ can again be seen as the local potential giving a density similar to the density of the fully nonlocal potential, to first order. The gap will, however, differ from the gap of the nonlocal potential, but, when corrected with the discontinuity
\be
\D_\xc=\bra c|\a\S_s^{\rm HF}-v^{\a}_{\x}| c \ket-\bra v|\a\S_s^{\rm HF}-v^{\a}_{\x}| v \ket
\label{discexx}
\ee
the gap is expected to be close to that of the nonlocal hybrid functional. Gaps calculated in this way using other exchange based functionals can be found in Refs. \onlinecite{kuisma,baerends}.

\subsection{Optimal $G_0W_0$ starting point based on a local hybrid potential}
The common crucial ingredient in $GW$, COHSEX and hybrid functionals is the nonlocal exchange term. Due to this similarity the hybrids can be used as a way to do approximate self-consistent $GW$.
Such an approach was developed in Refs. \onlinecite{atalla,opthyb2}. By using a hybrid as a starting-point for $G_0W_0$ the $\a$ parameter is varied until the $GW$ correction vanishes. At this value, the $GW$ and hybrid eigenvalues agree 
\be
\bra k|\S_s^{GW}(\ve^{nl}_{k})-\S_s^{\rm HYB,\a}| k \ket_{nl}=0\Rightarrow E^{GW}_k=\ve^{nl}_{k}.
\label{opt}
\ee 
Here the matrix elements are evaluated with orbitals generated by the nonlocal $\S_s^{\rm HYB,\a}$, emphasized by the sub(super)-script $nl$. This method has been shown to perform well for molecules, improving the ionization energies as compared to standard hybrid functionals and $G_0W_0$ based on the PBE starting-point.\cite{atalla} 
We note, however, that it is not possible to derive an equation similar to Eq.~(\ref{rpadisc}) combining the Klein $GW$ energy 
functional with a nonlocal potential. In fact, it has been shown to lack an extremum when varied in a restricted space 
of nonlocal but static potentials.\cite{sohrab10}

We will now present a variant that utilizes the RPA energy and, hence, the optimization with respect to a {\em local} potential.
As seen in the previous subsection a $G_0W_0$ correction based on the local RPA potential is justified via the $GW$ LSS equation (Eqs.~(\ref{lss}) and (\ref{gw1})). Analogously, a hybrid correction based on the local hybrid potential is justified via the hybrid LSS equation. We have
\be
E^{\rm HYB,\a}_k=\ve_{k}+\bra k|\S_s^{\rm HYB,\a}-v^{\rm hyb,\a}_{\rm xc}| k \ket.
\label{corrhyb}
\ee 

We will now approximate the RPA potential in Eq.~(\ref{gw1}) by the local hybrid potential
\be
E_k^{\rm RPA}\approx\ve_{k}+\bra k|\S^{GW}_s(\ve_{k})-v^{\rm hyb,\a}_{\rm xc}| k \ket
\label{rpalhyb}
\ee 
and optimize $\a$ such that the correction in Eq.~(\ref{corrhyb}) and Eq.~(\ref{rpalhyb}) are equal. This is equivalent to
\be
\bra k|\S^{GW}_s(\ve_{k})-\S_s^{\rm HYB,\a}|k\ket=0\Rightarrow E^{\rm RPA}_k=E^{\rm HYB,\a}_k,
\label{selfopt}
\ee
\begin{figure}[t]
\includegraphics[width=0.68\columnwidth,angle=90]{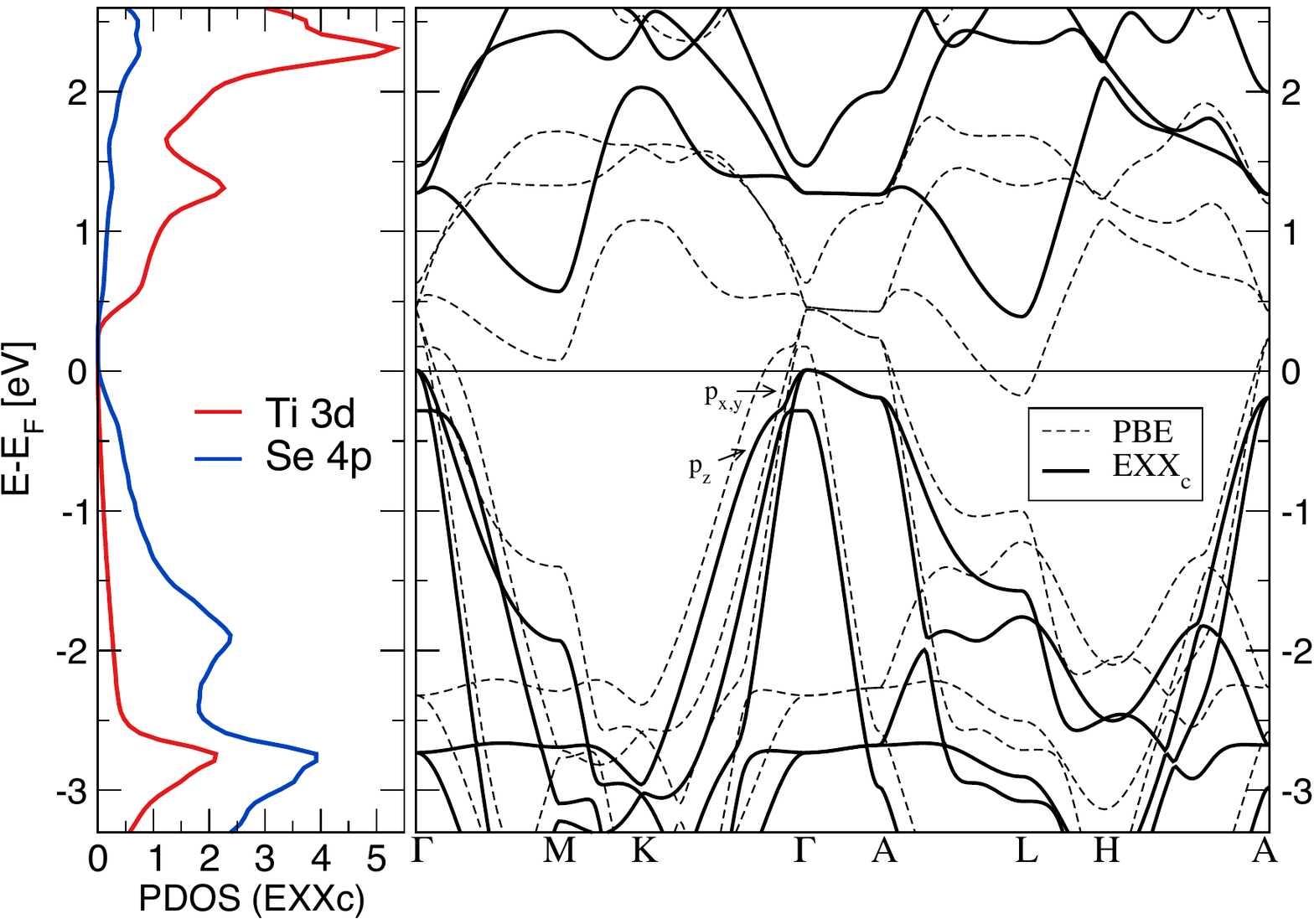}
\caption{Band structure and $p,d$-orbital projected density of states in the high-$T$ phase of TiSe$_2$. EXX$_c$ (full lines) compared to PBE (dashed lines). }
\label{pbevsexx}
\end{figure}
\begin{figure}[t]
\includegraphics[width=0.95\columnwidth,angle=0]{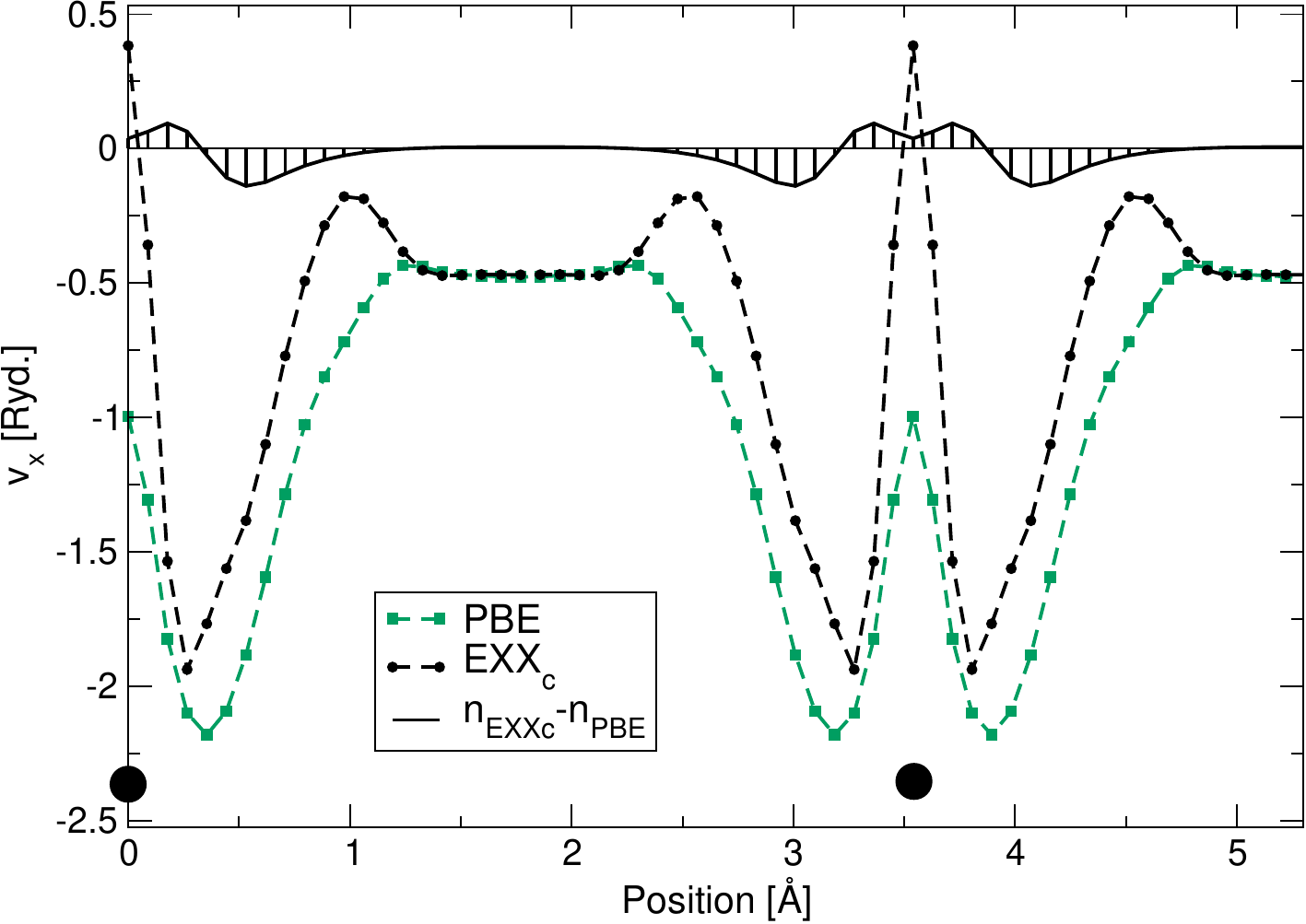}
\caption{Exchange part, $v_\x$, of the KS potential along the [100] direction. PBE in green with squares and EXX$_c$ in black with circles. 
The density difference generated by these potentials is also illustrated (with arbitrary unit on the y-axis). 
Black dots correspond to the position of the Ti-atoms.}
\label{tise2oep}
\end{figure} 
where the self-energy operators are evaluated with orbitals and eigenvalues from $v^{\rm hyb,\a}_{\rm xc}$  (instead of $\S_s^{\rm HYB,\a}$ as in Eq.~(\ref{opt}) above).
In this way, we optimize $\a$ such that the DFT hybrid functional behaves like the RPA functional when varying the particle number. 
The difference between this approach and the one in Ref.~\onlinecite{atalla} lies in which type of reference system is used to evaluate the GW energy. Allowing for nonlocal potentials can have a large impact on the energy due to the opening of a large gap.

Due to lack of frequency dependence in $\S_s^{\rm HYB,\a}$ the condition in Eq.~(\ref{selfopt}) is, in general, impossible to fulfil for all bands at every $k$, but can be made true for the difference between the highest occupied level and the lowest unoccupied level. One thus needs to optimize $\a$ using the following condition
\bea
\bra c|\S_s^{ GW}(\ve_{c})-\S_s^{\rm HYB,\a}| c \ket-\,\,\,\,\,\,\,\,\,\,\,\nn&&
\\\bra v|\S_s^{ GW}(\ve_{v})-\S_s^{\rm HYB,\a}| v \ket &=&0.
\label{lopt}
\eea
\begin{table*}[t]
\caption{\label{tab1} Band gaps (eV) of Ar, c-BN, ScN, TiS$_2$ and TiSe$_2$. The $G_0W_0$ results (without renormalization factor) are 
obtained with $G_0$ of the approximation in the preceding column. The H-$G_0W_0$ results are evaluated with the 
optimized nonlocal hybrid functional having an equal gap. The EXX results for Ar, BN and ScN are compared to EXX results found in the literature. EXX$_c$ corresponds to EXX 
plus a correlation term at the PBE level. All results (EXX, EXX$_c$ and lhyb) for TiS$_2$ and TiSe$_2$ are obtained with $\mu=0.1\,{\rm \AA}^{-1}$. 
}
\begin{ruledtabular}
\begin{tabular}{l r r r r r r r r r r r r r r r r r r r}
Solid  & \vline  & PBE  & $G_0W_0$& \vline& EXX & EXX lit.&EXX$_{\rm c}$&$G_0W_0$ &\vline & $\a$ & lhyb&  $G_0W_0$&\vline&$\a$&H-$G_0W_0$&\vline&HSE06&\vline&Exp.\\
\hline
Ar &\vline& 8.65	& 14.09 &\vline &9.57&9.61\footnote[1]{Reference \onlinecite{kresseoep}.}&9.93&14.30 &\vline& 0.57 & 9.38 & 14.19&\vline&0.63&14.88&\vline&10.32&\vline&14.2\footnote[2]{Reference \onlinecite{arexp}.}\\
BN &\vline&    4.54     & 6.51 &\vline & 5.58 &5.57\footnotemark[1]&5.12&6.78 &\vline&  0.25&4.70 &6.61&\vline&0.30&6.99&\vline &5.85&\vline&$6.4\pm0.5$\footnote[3]{Reference \onlinecite{bnexp}.}\\
ScN &\vline&   -0.05     & 1.01 &\vline & 1.57 & 1.58\footnote[4]{Reference \onlinecite{scnoep}.}&1.39&0.81&\vline&0.17&0.21 &0.96&\vline&0.24&1.51&\vline&0.92&\vline&$0.9\pm0.1$\footnote[5]{Reference \onlinecite{scnexp}.}\\
TiS$_2$ &\vline&   -0.10     &1.18 &\vline &  1.18&-& 1.03&0.30 &\vline&0.25&0.17 &0.82&\vline&0.33&1.17&\vline&0.57&\vline&$0.5\pm0.1$\footnote[6]{Reference \onlinecite{tis2exp}.}\\
TiSe$_2$ &\vline&  -0.63     & 0.37&\vline & 0.57 &-& 0.38&-0.85 &\vline&0.20&-0.45 &-0.07&\vline&0.32&0.40&\vline&-0.15&\vline&-0.1-0.1\footnote[7]{See text.}\\
\end{tabular}
\end{ruledtabular}
\end {table*}

Let us now compare this approach to other $GW$-schemes. The most common way to 
include some form of self-consistency within $GW$ is to iterate the eigenvalues, i.e., solving 
Eq. (\ref{gw_dft}), while keeping the KS orbitals fixed at the PBE level. This works well under the 
assumption that xc effects beyond PBE are
unimportant for the orbitals. An advantage of our approach 
is that it does not rely on this assumption as it takes 
into account both eigenvalues and orbitals at the same level of approximation. Another, more advanced 
approach, that takes into account changes in the orbitals via a static but nonlocal potential, is the 
"quasi-particle self-consistent $GW$" of Ref. \onlinecite{PhysRevLett.96.226402}. This approach requires, 
however, the generation of the full self-energy matrix and not just the diagonal terms, making it more 
expensive than the approach suggested here. Results from this approach have shown that, indeed, 
self-consistency in the orbitals is important.\cite{PhysRevB.82.115102}

In the next section we will show that self-consistency has a very small effect on systems where 
PBE already gives a good description of the orbitals. In contrast, for systems where exact-exchange plays 
an important role, self-consistency is necessary and we will show that, via  Eq.~(\ref{lopt}), 
it is possible to obtain meaningful results. The validity of the $G_0W_0$-approach is determined by 
the validity of the RPA for the given system. Furthermore, by approximating the RPA potential with the 
local hybrid potential we generate, as a bi-product, a hybrid functional that can be used to study 
other properties such as phonons and lattice instabilities. 

\section{Numerical results}
In this section we start by introducing TiSe$_2$ and the technical aspects of our calculations. We then 
present the $G_0W_0$ results for TiSe$_2$, TiS$_2$ and a set of well-known systems (Ar, c-BN and ScN) 
for which there already exist both EXX and $G_0W_0$ results in the literature. Finally, we investigate the 
performance of the RPA optimized hybrid functional in capturing the CDW phase of TiSe$_2$.

\subsection{System and computational details} 
The high-$T$ phase of TiSe$_2$ crystallizes in the space group $P\bar 3m1$. It belongs to the $1T$ family of the layered transition metal dichalcogenids with the Ti-atom octahedrally coordinated by six neighbouring Se-atoms. A semi-metallic behaviour is found in most experiments. Below 200 K a CDW transition occurs, characterised by a $2\times2\times2$ superstructure (space group $P3\bar{c}1$) and the opening of a small gap. The distortion pattern can be uniquely defined by the displacement $\delta$Ti and the ratio $\d$Ti$/\d$Se $\approx 3$. Standard DFT functionals predict a phonon instability at the three equivalent $L$ points. A symmetric combination of these gives the correct CDW pattern, but with a severely underestimated distortion amplitude.\cite{bianco2015} Hybrid functionals give a better description and have revealed the important role of HF exchange for the instability. This possibly hints to the presence of an excitonic instability. Although a weakly screened electron-hole interaction is clearly important, no spontaneous electronic CDW has so far been generated in bulk TiSe$_2$. Hybrid functionals induce the CDW via a strong electron-phonon coupling combined with the enhanced electronic susceptibility at the $L$ points. This mechanism is given support 
by the combined accuracy of the electronic bands, phonons and distortion amplitude.\cite{hellgrenbaima17}

In this work we aim for the more sophisticated $GW$ method that allows for a physical description of the screened interaction. Due to the increase in computational cost we have been limited to the high-$T$ phase. The low-$T$ CDW phase will be studied with the RPA-based hybrid functional, optimized according to the procedure described in the previous section. 

In addition, we have determined the gap of TiS$_2$ which is structurally identical to TiSe$_2$ but lacks a 
CDW transition, at least in the bulk. We have only found one $GW$ study of TiS$_2$ where the gap was determined 
to 0.75 eV,\cite{tis2gw} which can be compared to the experimental result of around 0.5 eV.\cite{tis2exp}
We have also looked at solid Ar (a van der Waals bonded large-gap insulator), c-BN (a $sp$ bonded insulator) and ScN (a $pd$ bonded semiconductor) 
in order to illustrate the workings and validity of the equations derived in Sec. II.

The hybrid calculations have been performed with VASP,\cite{vasp1,vasp2,vasp3} Quantum Espresso\cite{qe} (QE) and the {\rm CRYSTAL} program.\cite{doi:10.1002/wcms.1360}  Whenever comparisons could be made these codes, despite using different pseudopotentials, or in the latter case, a Gaussian basis-set, agree rather well. For example, $pd$ gaps agree within 0.05 eV. For TiS$_2$ and TiSe$_2$ we have 
used a range-separation parameter of $\mu=0.1\,{\rm \AA}^{-1}$ in all hybrid calculations.
We chose a range twice as large as in HSE06 since our previous work on TiSe$_2$ indicated that HSE06 was somewhat too short ranged.\cite{hellgrenbaima17} The local hybrid potential was calculated with QE using an iterative algorithm for solving the LSS integral equation (Eq. (\ref{exxlss})) (see Ref. \onlinecite{localrpa} for further details). The $GW$ self-energy was subsequently evaluated using the YAMBO code with full frequency integration.\cite{yambo1,yambo2} 
For testing the optimization scheme in Eq.~(\ref{opt}) we switched to the VASP code which allows the self-energy 
to be evaluated with a hybrid $G_0$. Agreement between different codes that use different numerical techniques 
and pseudopotentials is still hard to achieve within $GW$.\cite{PhysRevB.90.075125,RANGEL2020107242} Nevertheless, 
for Ar and c-BN, results on the PBE-$G_0W_0$ level agree within 0.05 eV. For the $pd$ gapped systems 
we found variations up to 0.1 eV (TiS$_2$ and TiSe$_2$) and 0.2 eV (ScN), which should be 
taken into account when comparing the different schemes. 
 We used a $12\times 12\times 6$ and $10\times 10\times 4$ $k$-point grid for TiSe$_2$ and TiS$_2$ respectively. Up to 
500 unoccupied bands where included in the self-energy. 
\begin{figure*}[t]
\includegraphics[scale=0.48]{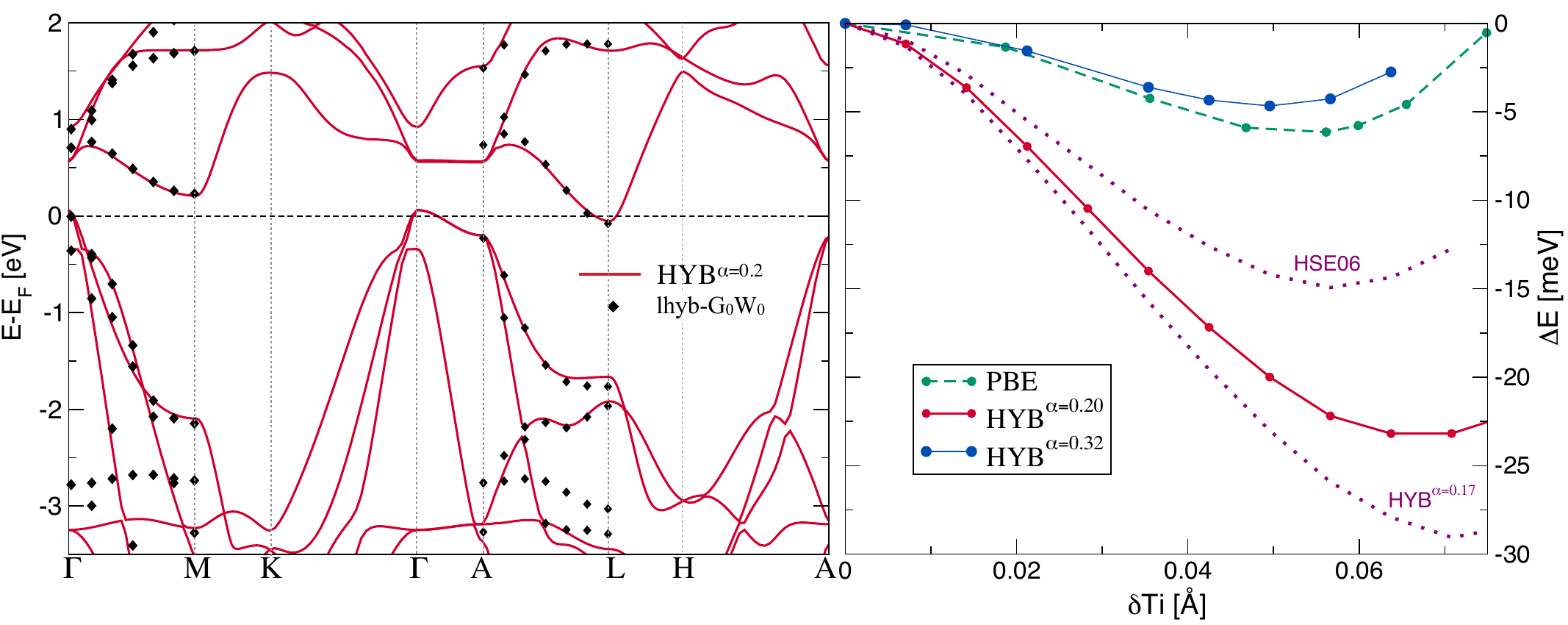}
\caption{Left: Band structure of the optimal hybrid functional with 20\% of nonlocal exchange (HYB$^{\a=0.2}$) compared to the corresponding lhyb-$G_0W_0$ result. Right: Energy gain in the supercell as a function of Ti-distortion keeping the ratio $\delta {\rm Ti}/\d {\rm Se}=3$ fixed. Optimal hybrid (red) is shown together with PBE (green) and the hybrid functional of the H-$G_0W_0$ optimization (blue, $\a=32\%$). We have also included HSE06 and HYB$^{\a=0.17}$($\mu=0.0\,{\rm \AA}^{-1}$) from Ref. \onlinecite{hellgrenbaima17}.}
\label{gwh20}
\end{figure*}
\begin{figure}[b]
\includegraphics[scale=0.45]{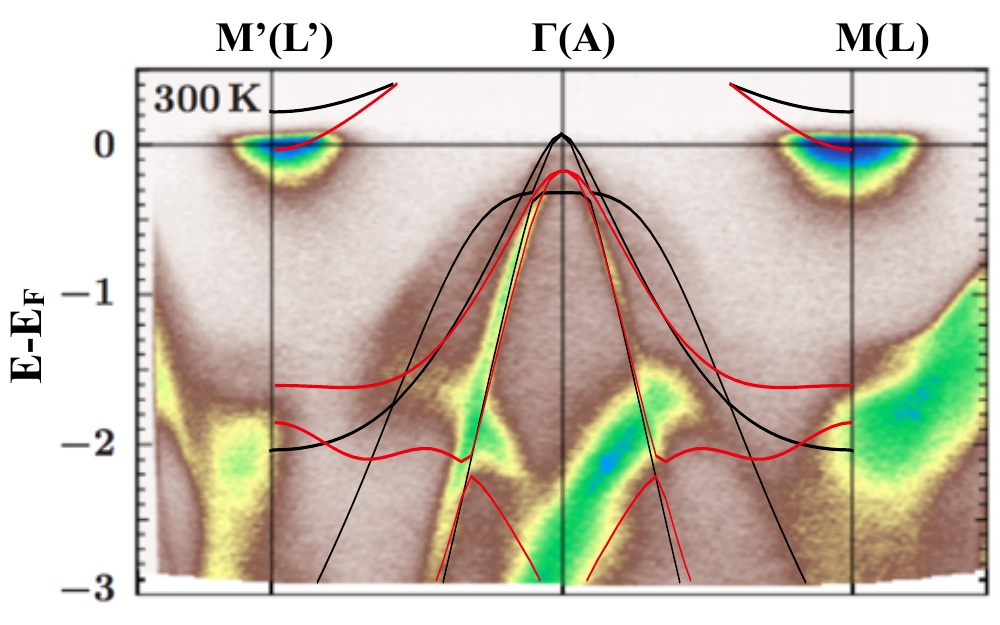}
\caption{Band structure of HYB$^{\a=0.2}$ compared to ARPES at 300 K.\cite{rohwer} Red lines corresponds to the $A-L$ path and black lines to the $\G-L$ path.}
\label{arpes}
\end{figure}
\subsection{$G_0W_0$ results}
In Table I we present the band gaps of Ar, c-BN, ScN, TiS$_2$ and TiSe$_2$. The PBE results are presented in the first column
and the $G_0W_0$ results, obtained on top of PBE orbitals, are presented in the second column. The renormalization factor 
is omitted in all $G_0W_0$ calculations. To demonstrate the accuracy of our implementation of the LSS equation, the gaps 
obtained within the EXX approximation (third column) are first compared to values found in the literature. We find a very good agreement in all 
cases where results are available. 
The EXX$_c$ results are obtained by adding the PBE correlation potential to the EXX potential. This EXX$_c$ potential is then used 
for obtaining the $G_0W_0$ results in the next column. In this way, we provide both extremes of the $\a$-range: PBE ($\a=0$) 
and EXX$_c$ ($\a=1.0$). The $\a$ parameter is then optimized according to Eq.~(\ref{lopt}). The optimized value of $\a$, 
the KS gap of the corresponding local hybrid potential (lhyb), and the final $G_0W_0$ gap are presented in the following 
three columns. As a result of Eq.~(\ref{lopt}), the $G_0W_0$ gap has to be the same as the perturbative gap of the 
nonlocal hybrid functional (Eq.~(\ref{corrhyb})) with parameter $\a$. 
Finally, we present the H-$G_0W_0$ results which are obtained using the optimization scheme of Eqs.~(\ref{opt}), the 
HSE06 results and experimental values.

Looking at the results for Ar and c-BN we immediately see that the $G_0W_0$ results are not so sensitive to which 
KS potential is used. The EXX potential increases the KS gap by around 1 eV but this leads only to a small 
increase of 0.2-0.3 eV in the $G_0W_0$ gap. By optimizing $\a$ we find a gap in between ($\a=0.57$ for 
Ar and $\a=0.25$ for c-BN). These values are consistent with RPA in a sense that both RPA and 
the hybrid functional give the same gap when evaluated with the orbitals of the optimized local hybrid potential. 
We note that the H-$G_0W_0$ results generally leads to larger values of $\a$. 

In ScN, a $pd$ semiconductor, we see a partially different behaviour. First, PBE predicts a semi-metallic ground state with a $pd$ band overlap. 
Including exact-exchange a KS gap of 1.57 eV opens. This rather large variation produces again only 
a small variation at the $G_0W_0$ level. However, the behaviour of the correction is opposite as compared to the 
correction in Ar and c-BN by giving a smaller $G_0W_0$ gap with EXX$_c$ than with PBE. This somewhat counterintuitive 
behaviour was noted already in Ref. \onlinecite{scn}. 
We now look at TiS$_2$ which also has a $pd$ gap. We see a similar trend but now the $GW$ variation is larger, 
ranging from 0.3 eV with EXX$_c$ to 1.18 eV with PBE. In this case, the optimization plays a crucial role. 
With 25\% of exchange the gap optimize to 0.82 eV. After this study we are now ready to turn to TiSe$_2$.

The high-$T$ phase PBE band structure has been published in several previous works, but is repeated here in Fig.~\ref{pbevsexx}. 
The PBE (and LDA) result differ strongly from ARPES experiments as analysed in detail in Ref. \onlinecite{bianco2015}. Similar to ScN and TiS$_2$ the Se-$p$ -Ti-$d$ band 
overlap is severely overestimated and, in this case, even inverted at $\G$, leading to strong $pd$ hybridization.\cite{bianco2015}
Furthermore, the $p_z$ orbitals corresponding to the flattened $p$-band around $\G$ is pushed above the Fermi level leading to 
excess $d$-electron occupation. These large errors invalidate the use of a standard PBE starting-point for $G_0W_0$. If we 
use PBE as a starting-point for $G_0W_0$, we open a gap of 0.37 eV between $\G$ and L (see Table I). The band gap is actually a bit smaller since 
we also found 'mexican hat' features around $\G$ similar to those reported in Ref. \onlinecite{cazzaniga2012}. 
These features are found already 
in the HF term but disappears as soon as the orbitals are updated.\cite{hellgrenbaima17} 

In Fig. \ref{pbevsexx} we also compare the KS band structure of the EXX$_c$ approximation to PBE. The corresponding projected density of 
states (PDOS) is shown in the side panel. 
The inclusion of exchange, even with a KS local potential, corrects the occupations and opens a gap
between $\G$ and $L$. Including the discontinuity (Eq.~(\ref{discexx})) the gap becomes as large as 3.75 eV, 
in agreement with a HF$_c$ calculation. 

In Fig. \ref{tise2oep} we have plotted the exchange part of the corresponding KS potentials (PBE and EXX$_c$). The accuracy of our 
LSS implementation can be seen from the smoothness of our EXX$_c$ potential. In EXX$_c$ we see additional structures around the 
Ti-atom that are missing in PBE. These barrier-like features typically act to localize charges. Looking at the density difference,
EXX$_c$ contracts the density around the Ti-atom. According to Eq.~(\ref{exxlss}) we expect a similar behaviour in the HF$_c$ approximation. 

If we evaluate $G_0W_0$ on top of the EXX$_c$ band structure the gap closes and we find a large band overlap (-0.85 eV). The magnitude of the variation is very close to the one in TiS$_2$ and it is clear that a self-consistent scheme 
is necessary. At optimal $\a=0.2$, we find a small band overlap of around 0.1 eV. ARPES has predicted results between -0.1 and 0.1 eV 
in the high-$T$ phase.\cite{PhysRevB.61.16213,PhysRevLett.88.226402,PhysRevB.65.235101,PhysRevLett.99.146403,PhysRevB.79.045116,chiang-arpes,aebi19} Our value is thus a reasonable prediction and shows that even a metallic solution can be found within $GW$. The electron-phonon mechanism found in Ref. \onlinecite{hellgrenbaima17} did not crucially depend on the existence of a Fermi surface, 
but a semi-metallic solution increases the probability for the existence of a purely electronic CDW. 

The band structure along $\G-M$ and $A-L$ is shown in Fig.~(\ref{gwh20}) superimposed on the full band structure of a hybrid functional 
with 20\% of exchange (HYB$^{\a=0.2}$). We see that not only the band overlap around the Fermi-level agrees but also the band dispersions. Dynamical effects in the self-energy seem important around -3 eV where the $pd$ mixed flat band is shifted downwards in the hybrid functional with respect to $G_0W_0$. Experiments place this band somewhere in between.\cite{monney3darpes} In Fig. (\ref{arpes}) we have superimposed the same results on an ARPES experiment by Rohwer et al.\cite{rohwer} Since spin-orbit coupling (SOC) is not included care should be taken when comparing with experiment. Previous studies have shown that SOC splits the degenerate $p$-bands at $\G$ which could have a small effect on the comparisons. Overall we see a very good agreement between theory and experiment noting that some of the deviations can be explained by looking at different values for $k_z$ (see discussion in Ref. \onlinecite{hellgrenbaima17}).
\begin{figure}[t]
\includegraphics[width=\columnwidth]{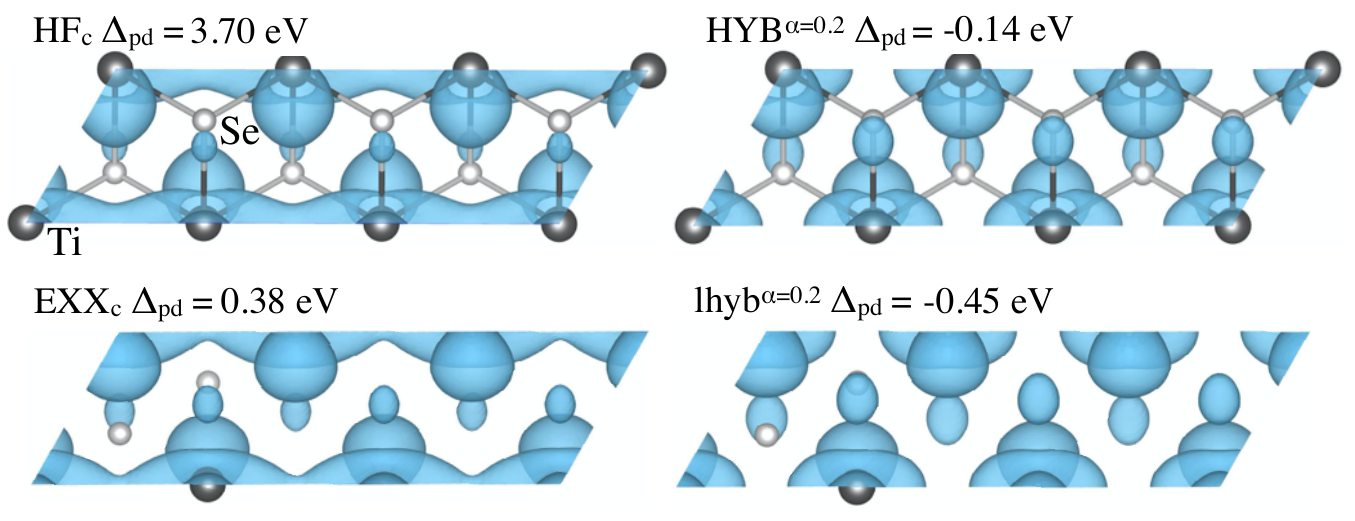}
\caption{Iso-surface of a Ti-$d$-orbital along the $\G-M$ path.\cite{vesta} To the left EXX$_c$ is compared to HF$_c$, and to the right the optimal hybrid with 20\% of nonlocal exchange is compared to the corresponding local hybrid approximation.}
\label{orbitals}
\end{figure}

To the right in Fig.~\ref{orbitals} we have compared an unoccupied $d$-orbital of the local hybrid with the same orbital of the nonlocal hybrid. The orbitals are very similar despite very different underlying gaps. The same is true for EXX$_c$/HF$_c$ to the left in the figure, suggesting that not only the density but also orbitals are well-mimicked by the local KS potential. The 
effect of exchange on the $d$-orbitals might be one explanation for the sensitivity of $G_0W_0$ to the fraction of exchange in the starting-point. The larger is $\a$ the more charge is localized between the Ti-atoms. The charge on the Se atoms, i.e., the hybridization with Se-$p$-orbitals, is instead seen to reduce with $\a$. The H-$G_0W_0$ yields a gap of 0.4 eV at 32\% of exchange, which is much larger than any experimental value. We also performed a self-consistent COHSEX calculation which gives a more reasonable result of 0.12 eV. 
We stress that these gaps are not related to the CDW since the symmetry is preserved in our calculations. 
\subsection{RPA optimized hybrid functional}
The approach applied above shows that $G_0W_0$ predicts a value for the $pd$ band-overlap which is in good agreement with many experiment. It also gives us a prediction for $\a$ based on the derivative of the RPA functional. This hybrid functional can now be used to study the CDW instability, too expensive for an approach like $GW$ or RPA. For an in-depth analysis of the CDW instability with hybrid functionals we refer the reader to Ref. \onlinecite{hellgrenbaima17}. Here we will restrict ourselves to a comparison between the hybrid functionals obtained via the different optimization procedures described in Sec. II.

To the right in Fig.~(\ref{gwh20}) we have used the RPA optimized $\a$ to calculate the energy gain in the supercell after distorting the atoms according to the CDW pattern. We have included both HSE06 and the 'best' hybrid functional ($\a=0.17,\mu=0.0\,{\rm \AA}^{-1}$) from Ref. \onlinecite{hellgrenbaima17} for comparisons. The set of parameters $\a=0.2,\mu=0.1\,{\rm \AA}^{-1}$ lies very close to the $(\a,\mu)$ path used in Ref. \onlinecite{hellgrenbaima17} and 
is not so different from the set that agreed best with experiment. Note, however, that in contrast to Ref. \onlinecite{hellgrenbaima17}, where the $\a$-parameter was fitted to the band gap and the phonon-frequencies, here it results from a self-consistent calculation.

If we look at the results for the H-$G_0W_0$ optimized hybrid functional with 32\% of exchange and a gap of 0.4 eV, we see that the 
energy gain strongly reduces. Both the energy gain and the $\d$Ti distortion worsen as compared to PBE. Given that the PBE underestimates the phonon mode associated with the CDW amplitude we do not expect this hybrid functional to accurately capture the CDW phase.
\section{Conclusions}
In this work we have applied a self-consistent $GW$ method to TiSe$_2$ in order to determine the much debated band-gap/band-overlap without adjustable parameters. We have also provided a theoretical justification for the choice of hybrid functional, i.e., the amount of admixture of exact exchange. 

First of all, it was found that the standard $G_0W_0$ prescription based on a PBE/LDA starting-point is unreliable due to qualitative errors in describing the band structure within PBE/LDA. To overcome this problem we have developed a simple quasi-self-consistent approach based on the local hybrid potential and the RPA functional. This approach allows for a systematic inclusion of exact-exchange in the starting-point, which, in the case of TiSe$_2$ has a large impact on, e.g., the description of the Ti-$d$-orbitals, and the resulting $pd$ gap. It is shown that $G_0W_0$ converges to a semi-metallic ground-state with a band-overlap of 0.1 eV. This is in line with transport experiments and many ARPES results, but contradicts first $G_0W_0$ results based on the LDA starting-point.

The $G_0W_0$ approach generates a hybrid $\a$-parameter consistent with RPA. With a motivated choice for $\mu$ this hybrid functional produces an electron-phonon coupling strong enough to induce the CDW transition. Furthermore, the potential energy surface lies very close to our earlier published hybrid results which gave very accurate phonons. While in our previous work, the $\a$-parameter was chosen as a best fit to both the band gap and the phonon frequencies, here it has been calculated via a self-consistent procedure involving the $G_0W_0$ method. The results are very similar, providing further support to the proposed mechanism of the CDW distortion in TiSe$_2$.

\acknowledgements
This work was performed using HPC resources from GENCI-TGCC/CINES/IDRIS (Grant No. A0050907625). 
M. H. and L. B. acknowledge support from Emergence-Ville de Paris. L. W. and M.C. acknowledge financial support from Agence Nationale de la Recherche (Grant N. ANR-19-CE24-0028)
and the Fond National de Recherche, Luxembourg via project INTER/19/ANR/13376969/ACCEPT.

\end{document}